\documentclass[12pt]{article}

\usepackage{graphicx}
\usepackage{amssymb}
\usepackage{amsmath}
\usepackage{color}
\usepackage{url}
\usepackage{graphicx}        % standard LaTeX graphics tool
\def\vrp{{\bf r}'}
\def\vr{{\bf r}}

\def\vn{{\bf n}}

\newcommand{\arxiv}{\let\arxivFig\matt}

\def\Xint#1{\mathchoice
   {\XXint\displaystyle\textstyle{#1}}%
   {\XXint\textstyle\scriptstyle{#1}}%
   {\XXint\scriptstyle\scriptscriptstyle{#1}}%
   {\XXint\scriptscriptstyle\scriptscriptstyle{#1}}%
   \!\int}
\def\XXint#1#2#3{{\setbox0=\hbox{$#1{#2#3}{\int}$}
     \vcenter{\hbox{$#2#3$}}\kern-.5\wd0}}

\def\dashint{\Xint-}

\graphicspath{{..//figures//}}

\title{Mathematical Analysis of the BIBEE Approximation for Molecular Solvation: Exact Results for Spherical Inclusions}
\author{Jaydeep P. Bardhan\\
Dept. of Molecular Biophysics and Physiology\\
Rush University Medical Center\\
Chicago IL 60612
\and
Matthew G. Knepley
Computation Institute\\
The University of Chicago\\
Chicago IL 60637}

\begin{document}
\maketitle

\begin{abstract}
We analyze the mathematically rigorous BIBEE (boundary-integral based
electrostatics estimation) approximation of the mixed-dielectric
continuum model of molecular electrostatics, using the analytically
solvable case of a spherical solute containing an arbitrary charge
distribution.  Our analysis, which builds on Kirkwood's solution using
spherical harmonics, clarifies important aspects of the approximation
and its relationship to Generalized Born models.  First, our results
suggest a new perspective for analyzing fast electrostatic models: the
separation of variables between material properties (the dielectric
constants) and geometry (the solute dielectric boundary and charge
distribution).  Second, we find that the eigenfunctions of the
reaction-potential operator are exactly preserved in the BIBEE model
for the sphere, which supports the use of this approximation for
analyzing charge-charge interactions in molecular binding.  Third, a
comparison of BIBEE to the recent GB$\epsilon$ theory suggests a
modified BIBEE model capable of predicting electrostatic solvation
free energies to within 4\% of a full numerical Poisson calculation.
This modified model leads to a projection-framework understanding of
BIBEE and suggests opportunities for future improvements.
\end{abstract}

\section{Introduction}
The strong influence of aqueous ionic solvent on biomolecular
structure and function necessitates its inclusion in almost all
theoretical studies in molecular
biophysics~\cite{Kirkwood34,Tanford57,Warshel84,Sharp90}, but for many
applications, including drug screening~\cite{Liu09} and protein
design~\cite{Harbury1998a,Vizcarra05}, explicit-solvent
MD~\cite{Brooks83,Case05,Phillips05} remains too expensive.
Implicit-solvent models\cite{Roux99,Cramer99,Simonson01,Baker05} offer
a computationally efficient alternative by approximating the solvent
influence using a potential of mean force (PMF) approach.  The most
popular models for the electrostatic component of this PMF are
continuum theories based on the Poisson or Poisson--Boltzmann partial
differential equations
(PDEs)~\cite{Kirkwood34,Tanford57,Warwicker82,BJBerne1,Honig7}.  For
all but the simplest molecular models, one must solve the PDE model
numerically, which requires substantial computational effort
regardless of whether one employs finite-difference
methods~\cite{Warwicker82,Gilson85,Klapper86,Gilson88,Nicholls1991a,Yu2007,Chen10,Cai10},
finite-element methods~\cite{You93,Cortis1997b,Baker01}, or the
boundary-element methods (BEM) based on boundary-integral equation
(BIE) reformulations of the
PDE~\cite{Miertus1981,Shaw85,Zauhar1985,Yoon90,Juffer91,Vorobjev1992,Zhou93,Purisima95,Horvath96,Liang97,Cances97,Lu05,Lu06,Bordner2003,Boschitsch02,Altman06,Altman09}.
The discrepancy between this computational effort, which calculates
the free energy due to solvent polarization, and the time to required
the other energies associated with a particular state (e.g., van der
Waals interactions) has motivated significant research towards
developing rapidly computed mathematical models that closely reproduce
the free-energy landscape of the Poisson equation.

Many of these fast models are designed to be integrated directly into
implicit-solvent MD, making differentiability of the energy function a
feature of crucial importance.  The generalized Born (GB) model
~\cite{Still90,Qiu97,Ghosh98,Dominy99,Onufriev00,Bashford00,Onufriev02,Grycuk03,Michel04,Romanov04,Sigalov05,Sigalov06,Tjong07,Mongan07,Mongan07_2}
is the most popular, but there are numerous others, notably the ACE
model of Schaefer and Karplus~\cite{Schaefer96}.  These approaches
introduce certain empirical parameters and analytical formulae, such
as effective Born radii in GB models, where the approximations rely on
physical insights into biomolecular electrostatics problems, including
the charge distributions, the dielectric constants, and the
near-spherical geometry of many globular proteins~\cite{Sigalov05}.
The ability of such models to capture broad features of the energy
landscape has led to numerous model refinements, parameterizations,
and implementations, but questions remain regarding these empirical
models' generality.

Such models contrast with the BIBEE approximate model we analyze in
this paper.  The BIBEE (boundary-integral based electrostatics
estimation) model derives from a systematic approximation of a
well-known BIE formulation of the mixed-dielectric Poisson
problem~\cite{Bardhan08_BIBEE,Bardhan09_bounds}, and represents a
complementary strategy to obtain an implicit-solvent model: a rigorous
operator approximation of the Poisson problem.  Important advantages
accrue by directly approximating the underlying operator problem
rather than exploiting specific features of biomolecular electrostatic
problems.  For example, the approximation can be analyzed
mathematically rather than empirically: we have shown that two
variants of BIBEE offer provable upper and lower bounds to the true
Poisson solvation free energy~\cite{Bardhan09_bounds}.  Furthermore,
there exists the possibility of developing equally rigorous
improvements to the approximation scheme.  The results in this paper
demonstrate both of these advantages and progress towards an accurate,
mathematically sound approximation scheme; however, it must be
acknowledged that the model and implementation are not yet suitable
for widespread adoption and application in MD.  The approach must
first overcome substantial challenges before it can be applied to
dynamical simulations in which the dielectric boundary changes at each
time step (a hurdle that has been surmounted only by a few developers
of advanced finite-difference methods, and not yet using
boundary-integral approaches).  In addition, an naive, unoptimized
implementation of BIBEE has been shown to be three to ten times slower
than a comparably unoptimized GB implementation\cite{Bardhan08_BIBEE},
a performance discrepancy of some importance as the community pursues
ever-longer dynamics simulations.

Here, we study the BIBEE model in the context of the analytically
solvable case of a spherical solute.  Using this model problem, we
highlight an interesting feature of most fast approximate
electrostatic models which has been studied in recent GB work and
reviews~\cite{Sigalov05}, but apparently never articulated explicitly:
most, but not all, fast methods assume that the solvation free energy
can be calculated using a separation of variables between the problem
geometry (here, the charge locations and dielectric boundary) and the
material properties, i.e., the dielectric constants.  This feature may
have implications for the design of improved implicit solvent models
or perhaps in other domains.  Although to our knowledge the
separability approximation has not been discussed directly in the
extensive GB literature~\cite{Bashford00,Baker05,Feig06}, it is
noteworthy that two of the most recent and accurate GB methods, the
GBMV (Generalized Born with Molecular Volume) model~\cite{Lee02} and
the modified GB$\epsilon$ of Sigalov and Onufriev \textit{et
  al.}~\cite{Sigalov05}, do incorporate corrections that explicitly
account for the errors inherent to the separability approximation.
The latter work, in particular, provides the basis for high accuracy
over a wide range of solute and solvent dielectric constants.
Furthermore, a recent review article highlights the material-dependent
corrections~\cite{Feig06}.

Analysis in spherical harmonics allows a straightforward proof that
for a sphere, the BIBEE approximation exactly preserves the
eigenfunctions of the Poisson reaction-potential operator.  The
operator eigenfunctions represent charge distributions that do not
interact via solvent polarization, and therefore a fast electrostatic
model should at least roughly capture eigenfunctions.  The sphere
geometry also permits a straightforward but detailed comparison of
BIBEE and the recently described GB$\epsilon$ model of Sigalov
\textit{et al.}, which is based on the most theoretically rigorous
analysis of the GB theory of which we are aware.  The mathematical
insights employed in deriving GB$\epsilon$ leads directly to two
improved BIBEE variants: one represents an new and tighter effective
lower bound, and another, much more accurate, version is based on the
central approximation in the GB$\epsilon$ model.  This accurate BIBEE
variant relies on a single fitting parameter and provides
electrostatic solvation free energies within a few percent of
numerical calculations.

The paper is organized as follows: the following section briefly
describes the Poisson electrostatic model under consideration,
Kirkwood's spherical-harmonics approach to deriving an closed-form
solution to the model~\cite{Kirkwood34}, and the BIBEE and
GB$\epsilon$ approximations.  Section~\ref{sec:separability} details
how a separability approach underlies most fast electrostatic schemes.
In Section~\ref{sec:analysis}, we employ the spherical harmonic
analysis to prove the BIBEE and continuum reaction-potential operators
share eigenfunctions.  In Section~\ref{sec:improved} we develop
improved BIBEE models using the GB$\epsilon$ analysis as a guide.
Section~\ref{sec:discussion} concludes the paper with a discussion of
implications, possible generalizations, and directions for future
work.

\section{Theory}

\subsection{Mixed-Dielectric Poisson Model and Boundary-Integral-Equation Formulation}
\begin{figure}[ht!]
\centering
\resizebox{3.0in}{!}{\includegraphics{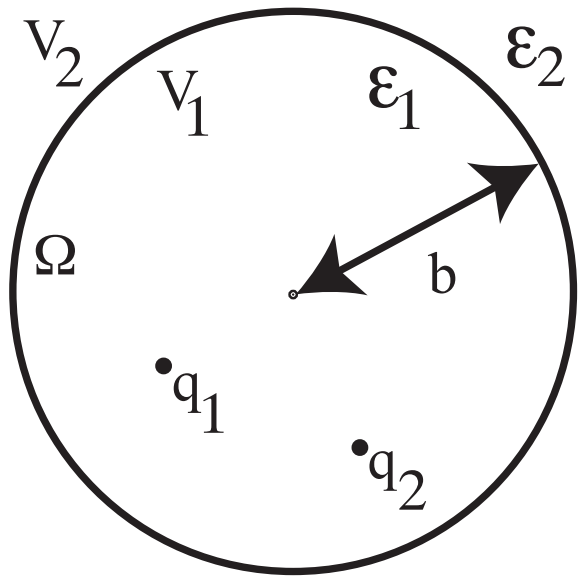}}
\caption{A diagram of the mixed-dielectric Poisson model, showing two
  discrete point charges in the
  solute.}\protect\label{fig:SphericalCavity}
\end{figure}
A diagram of our model is shown in Fig.~\ref{fig:SphericalCavity}.
The solute, labeled $V_1$, is a spherical cavity of radius $b$ with
boundary $\Omega$ and dielectric constant $\epsilon_1$.  The spherical
solute is embedded in an infinite homogeneous region, labeled $V_2$,
with dielectric constant $\epsilon_2$.  We assume without loss of
generality that the solute charge distribution consists of a set of
$Q$ point charges $\{q_k\}$ at locations $\vr_k$.  The electrostatic
potential in $V_1$ satisfies
\begin{equation}
  \nabla^2 \Phi_1(\vr) = -\frac{1}{\epsilon_1}\sum_{i=1}^Q q_k \delta(\vr-\vr_k)
\end{equation}
and the potential in $V_2$ satisfies the Laplace equation $\nabla^2
\Phi_2(\vr)=0$.  Across the boundary, the potential is continuous and
so is the normal component of the displacement field:
\begin{align}
  \Phi_1 |_{r=b}& = \Phi_2 |_{r=b}\label{eq:BCPot}\\
  \epsilon_1 \frac{\partial\Phi_1}{\partial r} |_{r=b} &= \epsilon_2 \frac{\partial\Phi_2}{\partial r} |_{r=b}\label{eq:BCNormalD}
\end{align}
where we have defined the normal direction pointing from $V_1$ into
the solvent region $V_2$.  Finally, the potential $V_2$ is assumed to
go to zero as $\vr\rightarrow\infty$.  The jump in dielectric
constants in the two regions gives rise to a discontinuity in the
polarization charge density as one crosses the dielectric
boundary~\cite{Jackson,Shaw85,Rush66,Boda04}.  This surface charge
density $\sigma(\vr)$ satisfies the boundary-integral equation
\begin{equation}
  \sigma(\vr) + \hat{\epsilon}\dashint_{\Omega} \frac{\partial}{\partial \vn(\vr)}\frac{1}{4\pi||\vr-\vrp||}\sigma(\vrp) d^2 \vrp
  =
  -\hat{\epsilon}\sum_{k=1}^{Q} q_k \frac{\partial}{\partial \vn(\vr)}\frac{1}{4 \pi ||\vr-\vr_k||}\label{eq:ASC}
  \end{equation}
where
\begin{equation}
\hat{\epsilon}=\frac{\epsilon_1-\epsilon_2}{\frac{1}{2}(\epsilon_1+\epsilon_2)}.
\end{equation}
and $\dashint$ denotes the Cauchy principal value integral.  In
operator notation, Eq.~(\ref{eq:ASC}) may be written
\begin{equation}
  \left(\mathcal{I}+\hat{\epsilon}\mathcal{D}^*\right)\sigma = B q\label{eq:defineB}
  \end{equation}
where $q$ is the vector of the fixed charges, $B$ maps $q$ to the
right-hand side of Eq.~(\ref{eq:ASC}), $\mathcal{I}$ is the identity
operator, and $\mathcal{D}^*$ is the normal electric-field operator.
The Coulomb potential induced by $\sigma(\vr)$,
\begin{equation}
  \psi(\vr) = \int_\Omega \sigma(\vrp)\frac{1}{4\pi||\vr-\vrp||}d^2 \vrp\label{eq:reac-pot},
\end{equation}
is the \textit{reaction-field potential}, and the electrostatic
component of the solvation free energy is
\begin{equation}
  \Delta G^{\mathrm{el}}_{solv} = \frac{1}{2} \sum_{k=1}^{Q}\psi(\vr_k)q_k,
  \end{equation}
Denoting the Coulomb integral operator of Eq.~(\ref{eq:reac-pot}) as $C$,
the linear mapping from the charge vector $q$ to the reaction
potential $\psi$, known as the \textit{reaction-potential operator}, is
\begin{equation}
  \psi = C \left(\mathcal{I}+\hat{\epsilon}\mathcal{D}^*\right)^{-1} B q
  \end{equation}
so the electrostatic solvation free energy is the quadratic form
\begin{equation}
  \Delta G^{\mathrm{el}}_{solv} = \frac{1}{2} q^T C \left(\mathcal{I}+\hat{\epsilon}\mathcal{D}^*\right)^{-1} B q.
  \end{equation}

\subsection{Kirkwood's Solution}
We review Kirkwood's series solution for our model
problem~\cite{Kirkwood34}. The potential inside $V_1$ is given by
\begin{equation}\label{eq:V1Pot}
  \Phi_1 = \sum^Q_{k=1} \frac{q_k}{\epsilon_1 \left|\vr - \vr_k\right|} + \psi,
\end{equation}
where again $\psi$ is the reaction potential. We expand $\psi$ in the
associated Legendre functions, keeping only terms valid within $V_1$, as
\begin{equation}\label{eq:ReactionPotentialExpansion}
  \psi = \sum^\infty_{n=0} \sum^n_{m=-n} B_{nm} r^n P_n^m(\cos\theta) e^{i m \phi}.
\end{equation}
The potential in region $V_2$ may be similarly expanded, keeping terms
valid at infinite $\vr$:
\begin{equation}\label{eq:Phi2Expansion}
  \Phi_2 = \sum^\infty_{n=0} \sum^n_{m=-n} \frac{C_{nm}}{r^{n+1}} P_n^m(\cos\theta) e^{i m \phi}.
\end{equation}
To determine the constants appearing in these expansions, we apply the
boundary conditions Eqs.~(\ref{eq:BCPot}) and~(\ref{eq:BCNormalD}) by
relating them to the Coulomb portion of $\Phi_1$, using the fact that
all charges are contained inside the sphere ($r_k < r$), as
\begin{eqnarray}
  \sum^Q_{k=1} \frac{q_k}{\epsilon_1 \left|\vr - \vr_k\right|} &=& \sum^Q_{k=1} \frac{q_k}{\epsilon_1}
    4\pi \sum^\infty_{n=0} \sum^n_{m=-n} \frac{1}{2n+1} \frac{r^n_k}{r^{n+1}} {Y_n^m}^*(\theta_k,\phi_k) Y_n^m(\theta,\phi) \\
  &=& \sum^Q_{k=1} \frac{q_k}{\epsilon_1}
    4\pi \sum^\infty_{n=0} \sum^n_{m=-n} \frac{1}{2n+1} \frac{r^n_k}{r^{n+1}}
    \sqrt{\frac{2n+1}{4\pi}\frac{(n-|m|)!}{(n+|m|)!}} \\
  && P_n^m(\cos\theta_k) e^{-i m \phi_k} \sqrt{\frac{2n+1}{4\pi}\frac{(n-|m|)!}{(n+|m|)!}} P_n^m(\cos\theta) e^{i m \phi} \\
  &=& \sum^\infty_{n=0} \sum^n_{m=-n} \frac{E_{nm}}{\epsilon_1 r^{n+1}} P_n^m(\cos\theta) e^{i m \phi},
\end{eqnarray}
where
\begin{equation}
  E_{nm} = \sum^Q_{k=1} q_k r^n_k \frac{(n-|m|)!}{(n+|m|)!} P_n^m(\cos\theta_k) e^{-i m \phi_k}.
\end{equation}
Now the first boundary condition, Eq.~(\ref{eq:BCPot}), gives us the
relation
\begin{equation}\label{eq:Cnm1}
  \frac{E_{nm}}{\epsilon_1} + b^{2n+1} B_{nm} = C_{nm}
\end{equation}
where we have equated each $(n,m)$ term in order for the it to hold
for all angles. In order to apply Eq.~(\ref{eq:BCNormalD}), we
differentiate each series term by term and equate them,
\begin{equation}\label{eq:Cnm2}
  \frac{1}{\epsilon_2} E_{nm} - \frac{\epsilon_1}{\epsilon_2} \frac{n}{n+1} b^{2n+1} B_{nm} = C_{nm}.
\end{equation}
We can eliminate the $C_{nm}$ coefficients, to give the reaction field
coefficients $B_{nm}$ in terms of the known source charge coefficients
$E_{nm}$,
\begin{equation}\label{eq:Bnm}
  B_{nm} = \frac{1}{\epsilon_1 b^{2n+1} } \frac{(\epsilon_1 - \epsilon_2) (n+1)}{\epsilon_1 n + \epsilon_2 (n+1)} E_{nm}.
\end{equation}

\subsection{BIBEE Approximations}
The BIBEE/CFA approximation replaces the boundary-integral operator of
Eq.~(\ref{eq:ASC}) with a scaled version of the identity operator, where
the scale factor is taken to be $-1/2$, which is the extremal
eigenvalue of the boundary-integral operator: thus we have
\begin{equation}
  \left(1-\frac{1}{2}\hat{\epsilon}\right)\sigma^{\mathrm{CFA}}= B q.\label{eq:bibee-cfa}
  \end{equation}
This eigenvalue is associated with the constant electric field at the
boundary~\cite{Bardhan08_BIBEE}, and is the reason that the CFA is exact for
a sphere with central charge; the CFA is actually exact for any charge
distribution that generates a constant electric field.  After
calculating an approximate surface charge distribution, the reaction
potential is computed just as if one had solved the actual boundary
integral equation of Eq.~(\ref{eq:ASC}), so that
\begin{equation}
  \Delta G^{\mathrm{CFA}} = \frac{1}{2} q^T C \left(1-\frac{1}{2}\hat{\epsilon}\right)^{-1} B q
  \end{equation}
The use of the extremal eigenvalue enables one to prove that the
BIBEE/CFA approximate electrostatic solvation free energy is an upper
bound to the actual electrostatic solvation free
energy~\cite{Bardhan09_bounds}.  The BIBEE/P approximation takes the
scale factor to be $0$, which is the other extremal eigenvalue for
spheres (and prolate spheroids):
\begin{equation}
  \sigma^{\mathrm{P}} = B q.\label{eq:bibee-p}
\end{equation}
The resulting approximate solvation free energy is a provable lower
bound for such surfaces. Though BIBEE/P is not a rigorous lower bound
for all surfaces, including oblate spheroids, tests on hundreds of
proteins showed that it never failed to provide an effective lower
bound~\cite{Bardhan09_bounds}.

Note that the approximate surface charge densities are scalar
multiples of one another, and thus the corresponding approximate
reaction-potential operators share a common eigenbasis.

\subsection{Generalized-Born Theory and the GB$\epsilon$ Model}
In the Generalized-Born (GB) electrostatic model, one associates with
each point charge an empirical parameter called an \textit{effective
  Born radius}, which is defined so that a spherical ion of that
radius would have the same solvation free energy as the original point
charge in the solute (i.e., the Born expression for the solvation
energy of a spherical ion).  In practice, one calculates the effective
Born radii approximately using e.g. the Coulomb-field approximation or
extensions thereof.  The reaction potential at a charge at $\vr_i$ due
to a unit charge at $\vr_j$, when the two effective radii are $R_i$
and $R_j$, and the distance between them is $r_{ij}$, is given by 
\begin{equation}
\varphi(\vr_i) = -\left(\frac{1}{\epsilon_1}-\frac{1}{\epsilon_2}\right)\frac{1}{f_{ij}^{\mathrm{GB}}(r_{ij},R_i, R_j)}
\end{equation}
where $f_{ij}^{\mathrm{GB}}$ is the usual Still equation
\begin{equation}
  f_{ij}^{\mathrm{GB}}(r_{ij},R_i,R_j) =
  \sqrt{r_{ij}^2 + R_i R_j\exp(-r_{ij}^2/4 R_i R_j)}.
  \end{equation}
The model has been demonstrated to exhibit remarkable fidelity to much
more expensive numerical solutions of the Poisson equation, but the
model's largely empirical nature poses challenges for extending the
model to more general configurations.  A substantial recent advance in GB theory was
achieved by Sigalov et al., who introduced the GB$\epsilon$ model to
provide improved accuracy of GB methods for a wider range of
dielectric constants~\cite{Sigalov05}, not just in the usual
biomolecular case in which $\epsilon_1/\epsilon_2 \ll 1$.

The GB$\epsilon$ model derives from analytically solvable cases of
point charges' self-energies in the limits $\epsilon_1/\epsilon_2 \to
\infty$ and $\epsilon_1/\epsilon_2 \to 0$ for a spherical solute.
Specifically, for an arbitrary solute, the authors define an
\textit{electrostatic radius} $A$ by computing
$\lim_{\epsilon_1/\epsilon_2\to \infty} \Delta G^{\mathrm{el}}_{ii} =
\Delta G_{ii}^{\mathrm{el}}(\epsilon_1/\epsilon_2\to\infty)$ and then
setting $A$ according to
\begin{equation}
  \Delta G^{\mathrm{el}}_{ii}(\epsilon_1/\epsilon_2 \to \infty)
  =
  -\frac{q_i^2}{2}\left(\frac{1}{\epsilon_1} -\frac{1}{\epsilon_2}\right)\frac{1}{A}\label{eq:es-radius-definition}.
  \end{equation}
The $i$th charge's effective Born radius for the limit
$\epsilon_1/\epsilon_2 \to 0$ is denoted by $\tilde{R}_i$; the slight
change of notation denotes the effective Born radius in the limit
$\epsilon_1/\epsilon_2 \to 0$, as opposed to the standard definition.
These parameters are defined by first computing each charge's
effective distance $r_i$ from the center of the molecule via
\begin{equation}
  \Delta G_{ii}^{\mathrm{el}}(\epsilon_1/\epsilon_2\to 0) = -\frac{q_i^2}{2}\left(\frac{1}{\epsilon_1}-\frac{1}{\epsilon_2}\right)\frac{1}{A-r_i^2/A}\label{eq:effective-position-definition}
\end{equation}
and then setting $\tilde{R}_i = A - r_i^2/A$ so as to recover the
familiar Born self-energy expression
\begin{equation}
  \Delta G_{ii}^{\mathrm{Born}}
  = -\frac{q_i^2}{2}\left(\frac{1}{\epsilon_1} -\frac{1}{\epsilon_2}\right)\frac{1}{\tilde{R}_i}
  \end{equation}
Kirkwood's exact expression for two charges' interaction in a sphere
of radius $A$ can be expressed succinctly via~\cite{Sigalov05}
\begin{equation}
 \Delta G^{\mathrm{el}}_{ij} = -q_i q_j \frac{1-\epsilon_1/\epsilon_2}{A \epsilon_1} \sum_{l=0}^{\infty} \frac{t_{ij}^l P_l(\cos\theta)}{1 + \frac{l}{l+1}\frac{\epsilon_1}{\epsilon_2}},
  \end{equation}
where $t_{ij} = |\vr_i| |\vr_j| / A^2$, $P_l(x)$ is the Legendre
polynomial, and $\theta$ is the angle between the charges with respect
to the origin. The GB$\epsilon$ model approximates this expression for
arbitrary molecular shapes, given the electrostatic radius $A$ and the
effective radii $\tilde{R}_i$, as
\begin{equation}
\Delta G^{\mathrm{el}}_{ij} =
-\frac{1}{2}\left(\frac{1}{\epsilon_1}-\frac{1}{\epsilon_2}\right) \frac{q_i q_j}{1+\alpha \epsilon_1/\epsilon_2}
\left[\frac{1}{f_{ij}^{\mathrm{GB}}(r_{ij},\tilde{R}_i,\tilde{R}_j)} + \frac{\alpha \epsilon_1/\epsilon_2}{A}\right]
\end{equation}
and the parameter $\alpha=0.57$ has been determined to minimize the
error between this approximation and the analytical
result~\cite{Sigalov05}.

Three important features of this model should be noted.  First, the
analysis of the GB$\epsilon$ model provides a mathematically justified
rationalization for the functional form of the Still equation, which
explains why GB methods enjoy the surprising success.  Second, the
method still requires the calculation (or estimation) of each charge's
self-energy, albeit in the limit $\epsilon_1/\epsilon_2 \to 0$ rather
than for the actual dielectric constants of interest.  Third, the
parameter $\alpha$ was determined by first expressing the pairwise
interaction between charges as a sum over spherical harmonics and then
manipulating terms to develop an accurate approximation to the
infinite sum.

\section{Separability}\label{sec:separability}
In both the BIBEE and standard GB approaches, the reaction potential at
$\vr_i$ due to a +1$e$ charge at $\vr_j$ can be written in the general
form
\begin{align}
  \psi(\vr_i) = g(\epsilon_1,\epsilon_2) h(V_1, \vr_i, \vr_j)\label{eq:separable}
  \end{align}
where $g(\epsilon_1,\epsilon_2)$ is a function only of the dielectric
constants and $h(V_1, \vr_i, \vr_j)$ is a function only of the solute
geometry (denoted by the volume) and the charge positions.  Thus, both
fast electrostatic models employ a \textit{separable} functional form
for the reaction potential.  Ample evidence supports that these
methods exhibit remarkable properties in capturing the electrostatic
free energy of solvation, but surprisingly there exists no clear
argument why a separable representation might be accurate: that is,
for the Poisson equation nothing in either the partial-differential
form
\begin{equation}
  \nabla \cdot \left(\epsilon(\vr)\nabla\varphi(\vr)\right) = -\rho(\vr)
  \end{equation}
or its boundary-integral form would \textit{a priori} suggest that one
could obtain an adequate approximation by moving to a separable
representation.

Formally, and specializing the analysis to the sphere case, an
off-center charge excites at least one mode for each multipole order;
as shown by Sigalov \textit{et al.}, the excitation/response relation
for a mode depends on both the multipole order $l$ and the dielectric
ratio $\beta = \epsilon_2/\epsilon_1$ via
\begin{equation}
  \frac{1}{1 + \beta \frac{l}{l+1}},
  \end{equation}
and therefore it is not possible to define an exact separable energy
function.  The many successes of GB theory clearly argue that
separable energy functions provide satisfactory accuracy for many
applications, and two more arguments can also be made to support their
study.  First, improvements on the original GB/CFA approach have
almost exclusively modified the function $h$, and in particular the
formulae associated with calculating the effective Born radii; thus,
the observed substantial advances in GB models have virtually all
maintained separability as a central mathematical feature.  Second,
earlier studies of BIBEE illustrated that modern GB methods very
accurately capture the eigenvalues of the reaction-potential operator
and reproduce the operator eigenvectors only with moderate accuracy;
in contrast, BIBEE models provide an excellent approximation to the
eigenvectors but only moderate reproduction of the eigenvalues.
Because these two drastically different approximations can capture
these distinct features with only minor modifications to separability,
it seems quite possible that one approximation may be found that
captures the best features of both approaches.

In GB theory, deviations from separable response are compensated
somewhat by the fact that the radius $R_i$ is set so that even if the
potential itself is large, the error in $\psi^{\mathrm{GB}}(\vr_i)$
is zero (assuming perfect radii).  Because the Still equation is a
smooth function designed to approximately capture the
distance-dependence of the reaction potential~\cite{Still90}, the
error will be relatively small nearby as well, i.e. for the closest
charges.  This interpolatory feature is an extra source of accuracy
for GB-type methods in standard situations with $\epsilon_1 <<
\epsilon_2$; nevertheless, as pointed out by Onufriev and
collaborators, high accuracy is not a general property for all values
of these two parameters.

Sigalov \textit{et al.} presented two key insights that provide a
strong theoretical basis for designing accurate and rapidly computed,
but not separable methods such as their GB$\epsilon$ model.  First,
the lowest mode $l=0$ is associated with the response due to the
solute monopole moment.  This mode is distinct because it is
associated with a constant potential inside, which is why GB$\epsilon$
makes effective use of the seemingly counterintuitive limit of a
conductor solute (i.e. $\epsilon_1 \to \infty$): a conductor has a
constant potential inside it.  Treating this mode distinctly, as
GB$\epsilon$ does, offers certain advantages and we return to this
idea later to improve BIBEE.

The next crucial insight in the GB$\epsilon$ model is that by
separating the $l=0$ term, it becomes much more reasonable to capture
a reasonable approximation to the relationship between the dielectric
constants and the ratio $l/(l+1)$.  For the higher order terms $l = 1,
2, \ldots$, the mode/dielectric ratio coupling terms are all between
$\frac{1}{2}\beta$ and $\beta$, and the highest frequency modes
generally exert a very small influence on the total energetics.  In
constructing the GB$\epsilon$ model, the authors minimized the error
in the solute reaction potential and found that $\frac{l}{l+1} \approx
0.57$ (this is the parameter $\alpha$) sufficed to give excellent
accuracy with a purely separable representation.  This value lies
between the dipole and quadrupole terms ($l=1$ and $l=2$).

\section{Analysis of BIBEE for a Spherical Solute}\label{sec:analysis}
The continuity of the normal dielectric displacement can also be interpreted as the accumulation of a single-layer
surface charge $\sigma$,
\begin{eqnarray}
  \sigma &=& \left[ E_\perp \right] \\
         &=& \frac{\partial\Phi_2}{\partial r} - \frac{\partial\Phi_1}{\partial r} \label{eq:sigma} \\
         &=& \frac{\epsilon_1}{\epsilon_2} \frac{\partial\Phi_1}{\partial r} - \frac{\partial\Phi_1}{\partial r} \\
         &=& \frac{\epsilon_1 - \epsilon_2}{\epsilon_2} \frac{\partial\Phi_1}{\partial r}
\end{eqnarray}
where the third line makes use of Eq.~(\ref{eq:BCNormalD}). In the
original formulation of the BIBEE approximation, we replace the
integral operator mapping the normal electric field at the surface to
a charge density with a diagonal
approximation~\cite{Bardhan08_BIBEE}. However, one may alternatively
consider this approximation as a deformation of the above boundary
condition to the requirement that the induced surface charges be
proportional to the Coulomb Field Approximation (CFA), i.e. neglecting
the reaction-field component:
\begin{equation}\label{eq:CFAsigma}
  \sigma^{CFA} = \frac{\epsilon_1 - \epsilon_2}{\epsilon_2} \frac{\partial\Phi^C_1}{\partial r}
\end{equation}
where
\begin{equation}\label{eq:Phi1C}
  \Phi^C_1 = \Phi_1 - \psi = \sum^Q_{k=1} \frac{q_k}{\epsilon_1 \left|\vr - \vr_k\right|}.
\end{equation}
Using Eqs.~(\ref{eq:CFAsigma}) and~(\ref{eq:sigma}), we can derive a boundary condition for the BIBEE/CFA approximation
\begin{equation}\label{eq:BCCFA}
  \frac{\epsilon_1}{\epsilon_2} \frac{\partial\Phi^C_1}{\partial r} = \frac{\partial\Phi_2}{\partial r} - \frac{\partial\psi}{\partial r}.
\end{equation}
We can now derive a similar relation to Eq.~(\ref{eq:Cnm2}) by equating coefficients,
\begin{equation}\label{eq:Cnm2CFA}
  \frac{1}{\epsilon_2} E_{nm} - \frac{n}{n+1} b^{2n+1} B_{nm} = C_{nm}
\end{equation}
which using Eq.~(\ref{eq:Cnm1}) gives
\begin{equation}\label{eq:BnmCFA}
  B^{CFA}_{nm} = \frac{\epsilon_1 - \epsilon_2}{\epsilon_1 \epsilon_2} \frac{n+1}{2n+1} \frac{1}{b^{2n+1}} E_{nm}.
\end{equation}
We emphasize that Eq.~(\ref{eq:BnmCFA}) shows explicitly that the
approximation has produced a separated representation in terms of $n$
and $\epsilon$.  We can derive a similar expression for the
approximate lower bound BIBEE/P given that
\begin{equation}\label{eq:Psigma}
  \sigma^P = \frac{\epsilon_1 - \epsilon_2}{\frac{1}{2} (\epsilon_1 +\epsilon_2)} \frac{\partial\Phi^C_1}{\partial r},
\end{equation}
which leads to the modified boundary condition
\begin{equation}\label{eq:BCP}
  \frac{3\epsilon_1 - \epsilon_2}{\epsilon_1 + \epsilon_2} \frac{\partial\Phi^C_1}{\partial r} = \frac{\partial\Phi_2}{\partial r} - \frac{\partial\psi}{\partial r}.
\end{equation}
We can now derive a relation analogous to both Eqs.~(\ref{eq:Cnm2}) and~(\ref{eq:Cnm2CFA}) by equating coefficients,
\begin{equation}\label{eq:Cnm2P}
  \frac{3\epsilon_1 - \epsilon_2}{\epsilon_1 (\epsilon_1 + \epsilon_2)} E_{nm} - \frac{n}{n+1} b^{2n+1} B_{nm} = C_{nm}
\end{equation}
which using Eq.~(\ref{eq:Cnm1}) gives
\begin{equation}\label{eq:BnmP}
  B^{P}_{nm} = \frac{2 (\epsilon_1 - \epsilon_2)}{\epsilon_1 (\epsilon_1 + \epsilon_2)} \frac{n+1}{2n+1} \frac{1}{b^{2n+1}} E_{nm}.
\end{equation}
Removing the common factor from Eqs.~(\ref{eq:Bnm}),~(\ref{eq:BnmCFA}), and~(\ref{eq:BnmP}),
\begin{equation}
  \gamma_{nm} = \frac{\epsilon_1 - \epsilon_2}{\epsilon_1} (n+1) \frac{1}{b^{2n+1}} E_{nm},
\end{equation}
so that
\begin{eqnarray}
  B_{nm}      &=& \frac{1}{\epsilon_1 n + \epsilon_2 (n+1)} \gamma_{nm} \label{eq:BnmDef} \\
  B^{CFA}_{nm} &=& \frac{1}{\epsilon_2} \frac{1}{2n+1} \gamma_{nm}  \label{eq:BnmCFADef}\\
  B^{P}_{nm}  &=& \frac{1}{\epsilon_1 + \epsilon_2} \frac{1}{n+\frac{1}{2}} \gamma_{nm}  \label{eq:BnmPDef}.
\end{eqnarray}
Clearly, if $\epsilon_1 = \epsilon_2 = \epsilon$, i.e. in a uniform medium with no reaction term $\psi$, then
both approximations are exact:
\begin{equation}
  B_{nm} = B^{CFA}_{nm} = B^{P}_{nm} = \frac{1}{\epsilon (2n+1)} \gamma_{nm}.
\end{equation}

\subsection{BIBEE Reaction-Potential Eigenfunctions Are Exact}
We can now show that the approximate BIBEE reaction-potential
operators have identical eigenspaces to the original operator, by
examining the effect on a unit spherical harmonic charge distribution
input
\begin{equation}
  \hat \rho_{n'm'} = P_{n'}^{m'}(\cos\hat\theta) e^{-i m' \hat\phi}.
\end{equation}
This source produces a Coulomb potential $\hat\Phi^C_1$ whose
expansion is defined by
\begin{equation}
  \hat E_{nm} = \hat r^n P_n^m(\cos\hat\theta) e^{-i m \hat\phi} \delta_{nn'} \delta_{mm'}
\end{equation}
where $\delta_{ij}$ is the Kronecker delta.  Then, from Eqs.~(\ref{eq:BnmDef})--(\ref{eq:BnmPDef}), it is clear that
\begin{eqnarray}
  \hat B_{nm} &=& \frac{(\epsilon_1 - \epsilon_2) (n+1)}{\epsilon_1 \left(\epsilon_1 n + \epsilon_2 (n+1)\right)} \frac{\hat r^n}{b^{2n+1}} P_n^m(\cos\hat\theta) e^{-i m \hat\phi} \delta_{nn'} \delta_{mm'}\\
  \hat B^{CFA}_{nm} &=& \frac{(\epsilon_1 - \epsilon_2) (n+1)}{\epsilon_1 \epsilon_2 (2n+1)} \frac{\hat r^n}{b^{2n+1}} P_n^m(\cos\hat\theta) e^{-i m \hat\phi} \delta_{nn'} \delta_{mm'}\\
  \hat B^{P}_{nm} &=& \frac{(\epsilon_1 - \epsilon_2) (n+1)}{\epsilon_1 \left(\epsilon_1 + \epsilon_2\right) \left(n+\frac{1}{2}\right)} \frac{\hat r^n}{b^{2n+1}} P_n^m(\cos\hat\theta) e^{-i m \hat\phi} \delta_{nn'} \delta_{mm'}
\end{eqnarray}
so that each reaction-field has a response only in the input harmonic
$(n',m')$. Therefore each input $\hat\rho_{n'm'}$ is an eigenfunction
of the exact reaction-potential operator and also an eigenfunction of
the BIBEE operators. Because the spherical harmonics form a complete
basis, the eigenspaces are identical.

%  \subsection{Errors}
%
%We may look at the deviation of our estimates from the exact expansion. First, for BIBEE/P, we have
%\begin{equation}
%  \frac{B_{nm} - B^{P}_{nm}}{B_{nm}} = \frac{1}{2n+1} \frac{\epsilon_1 - \epsilon_2}{\epsilon_1 + \epsilon_2}.
%\end{equation}
%Using Parseval's Theorem, the $L_2$ norm of the error in the reaction potential on the surface of the unit sphere is
%given by the error in the expansion coefficients,
%\begin{equation}
%  || \psi_1 - \psi^{P}_1 ||^2 = \sum^\infty_{n=0} \sum^n_{m=-n} (B_{nm}- B^{P}_{nm})^2 \frac{\epsilon_1 - \epsilon_2}{\epsilon_1} (n+1) E_{nm} \frac{4\pi}{2n+1} \frac{(n-|m|)!}{(n+|m|)!}
%\end{equation}

\subsection{Asymptotic Behavior of BIBEE Approximations}\label{subsec:Asymptotics}
Comparing Eqs.~(\ref{eq:Bnm}) with~(\ref{eq:BnmCFA}), we see that the $n =
0$ mode is exact in the BIBEE/CFA approximation,
\begin{equation}
  B_{00} = B^{CFA}_{00} = \frac{\gamma_{00}}{\epsilon_2},
\end{equation}
whereas BIBEE/P approaches the exact response in the limit $n\to\infty$:
\begin{equation}
  \lim_{n\to\infty} B_{nm} = \lim_{n\to\infty} B^{P}_{nm} = \frac{1}{(\epsilon_1 + \epsilon_2) n} \gamma_{nm}.
\end{equation}
In most biomolecule modeling problems, the cavity $V_1$ has a very low
dielectric constant compared to the surrounding medium $V_2$ (i.e. if
$\epsilon_1 \ll \epsilon_2$).   In the limit $\epsilon_1/\epsilon_2 \to 0$,
\begin{eqnarray}
\lim_{\epsilon_1/\epsilon_2\to 0}  B_{nm}      &=& \frac{\gamma_{nm}}{\epsilon_2 (n+1)} \label{eq:LargeEpsilonDiff} \\
\lim_{\epsilon_1/\epsilon_2\to 0}  B^{CFA}_{nm} &=& \frac{\gamma_{nm}}{\epsilon_2 (2n+1)},\\
\lim_{\epsilon_1/\epsilon_2\to 0}  B^{P}_{nm}  &=& \frac{\gamma_{nm}}{\epsilon_2 \left(n + \frac{1}{2} \right)},
\end{eqnarray}
so that the approximation ratios are
\begin{eqnarray}
  \frac{B^{CFA}_{nm}}{B_{nm}} &=& \frac{n+1}{2n+1}, \\
  \frac{B^{P}_{nm}}{B_{nm}}   &=& \frac{n+1}{n + \frac{1}{2}}.
\end{eqnarray}
We see that BIBEE/CFA is exact in the case of the uniform field, and
the modal contribution can be off by a factor of 2 for very high
spatial-frequency charge distributions. This mirrors the bounds
derived previously~\cite{Bardhan08_BIBEE}. In the case of BIBEE/P, the
situation is reversed: the uniform field can be incorrect by a factor
of 2, whereas the high frequency field is exact in the
limit. Furthermore, it is clear that BIBEE/CFA underestimates
coefficients, which is why BIBEE/CFA solvation free energy estimates
are rigorous upper bounds for the true solvation free energies (which
are negative quantities)~\cite{Bardhan09_bounds}.  Conversely, BIBEE/P
overestimates the coefficients and give lower bounds.

These asymptotic relations can be compared to Grycuk's
analysis~\cite{Grycuk03}. The BIBEE/CFA approximation has an exact
monopole moment, corresponding to Grycuk's case of a charge centered
in region $V_1$. Likewise, the observed inaccuracies for charges near
the cavity surface correspond to higher multipole moments, where we
see that BIBEE/CFA overestimates the coefficients by a factor of 2,
the same factor found by Grycuk.  The BIBEE/P model easily provides an
accurate approximation in this limit.

\section{Improved BIBEE Models}\label{sec:improved}
The considerations in Section~\ref{sec:separability} suggest a
strategy to improve the BIBEE model: one should use the BIBEE/CFA
approximation for the monopole moment and a different approximation
for the higher-order terms.  We explore two approximations here.  In
the simplest, BIBEE/P is used for the other terms; in the second, we
follow the approach in GB$\epsilon$ and use an effective parameter.
By using BIBEE/CFA only for the monopole moment, one ensures that the
constant component of the reaction potential is captured exactly, just
as the GB$\epsilon$ model was designed to do.  The strategy is easy to
adopt as an improved BIBEE model because it is associated with the
constant surface-charge distribution, and therefore the separation is
analytically exact.

For the sphere, the reaction potential expansion in this hybrid model
is given by $B^{CFA}_{00}$ in combination with $B^P_{nm}$ for $n, m
\ne 0$.  The alternate interpretation, suitable for application to
general geometries, is that one calculates an approximate surface
charge $\hat{\sigma}$ as a sum of BIBEE/CFA and BIBEE/P components:
\begin{equation}
  \hat{\sigma} = \sigma^{\mathrm{CFA,monopole}} + \sigma^{\mathrm{P,other}}.
  \end{equation}
To obtain these components, one first computes $B q$ as before,
i.e. the Coulomb electric field at the boundary.  This field is then
decomposed into two terms: its mean value $<B q>$, and the field minus
the mean, $B q - <B q>$.  One then employs the usual approximations
(\ref{eq:bibee-cfa}) and (\ref{eq:bibee-p}), so that
\begin{eqnarray}
(1 - \frac{1}{2}\hat{\epsilon}) \sigma^{\mathrm{CFA,monopole}} &=& <B q>\\
\sigma^{\mathrm{P,other}} &= &B q - <B q>.
  \end{eqnarray}
This approach has been validated using a sphere with random charge
distributions, and we note that this combination of approximations
provides a correction for charged species but gives exactly the
BIBEE/P result for net-neutral charge distributions.

Note that the modified BIBEE approach (BIBEE/M) leads to a reaction
potential that is not separable in the sense that it may be written in
the form of Eq.~(\ref{eq:separable}); instead, the potential is the
sum of two terms that are themselves separable.  Defining the operator
$B$ of Eq.~(\ref{eq:defineB}) via $B = \hat{\epsilon}\tilde{B}$, the
BIBEE/M reaction potential is
\begin{eqnarray}
  \psi^M &=& C \hat{\sigma} \\
         &=& \hat{\epsilon}C\tilde{B}q + \hat{\epsilon}\left((1-\frac{1}{2}\hat{\epsilon})^{-1}-1\right)C<\tilde{B}q>.
  \end{eqnarray}

\begin{figure}[ht!]
\centering
\resizebox{6.5in}{!}{\includegraphics{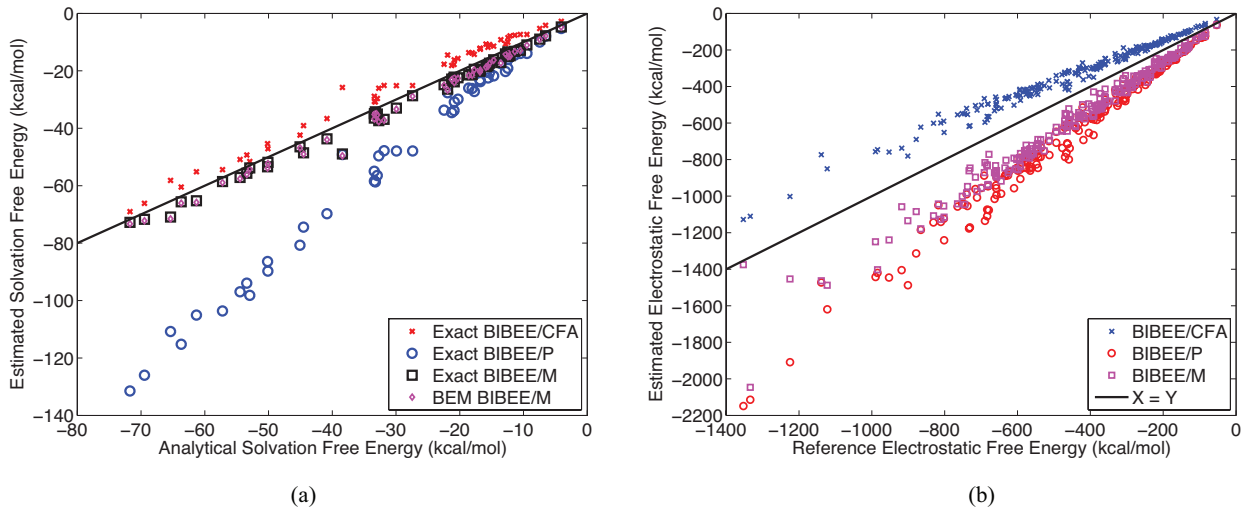}}
\caption{Modifying BIBEE using BIBEE/CFA for the monopole moment and
  BIBEE/P provides a tighter lower bound for charged molecules. (a)
  Correlation plot of the BIBEE estimated electrostatic solvation free
  energies, for multiple sets of 25 randomly located point charges in
  a sphere of radius 5~\AA, with dielectric constants $\epsilon_1=4$
  and $\epsilon_2=80$. Results labeled BEM BIBEE/M illustrate the same
  model computed using the BEM implementation of BIBEE, i.e. a
  numerical method suitable for arbitrary geometries. (b) Correlation
  plot of the improved BIBEE estimates for electrostatic solvation
  free energies for 200 proteins from the test set of Feig et
  al~\cite{Feig04_GB_vs_PB}, using CHARMM22 radii and charges and
  dielectric constants $\epsilon_1=4$ and
  $\epsilon_2=80$.}\protect\label{fig:sphere-correction}
\end{figure}

Figure~\ref{fig:sphere-correction}(a) shows exact and approximate
electrostatic solvation free energies for sets of 25 randomly located
point charges in a sphere of radius 5~\AA, where the point charges are
randomly assigned values less than 0.5 in magnitude; the dielectric
constants are taken to be $\epsilon_1 = 4$ and $\epsilon_2 = 80$.  As
expected, this approach is always a more accurate estimate than
BIBEE/P; in fact, because the monopole mode is captured exactly and
the others are overestimated when $\lambda = 0$, BIBEE/M is a tighter
lower bound for the sphere than BIBEE/P.  Though the majority of
random charge distributions see significantly improved estimates
between BIBEE/P and BIBEE/M, contributions from other modes can still
strongly affect the overall free energy and lead to little
improvement.  However, employing the method on 200 proteins from the
Feig \textit{et al.}  test set~\cite{Feig04_GB_vs_PB} illustrates that
the improvement is quite modest at best in practice, as shown in
Figure~\ref{fig:sphere-correction}(b).  These calculations were
conducted using CHARMM22 radii and charges~\cite{MacKerell98},
$\epsilon_1=4$, $\epsilon_2=80$, and the molecular surface with probe
radius 1.4~\AA; triangular discretizations were computed using
MSMS~\cite{Sanner96}.  Clearly, despite improvements in accuracy, more
improvement is still needed in order to predict solvation free
energies with the same accuracy as sophisticated Generalized Born
methods.

\begin{figure}[ht!]
\centering
\resizebox{3.0in}{!}{\includegraphics{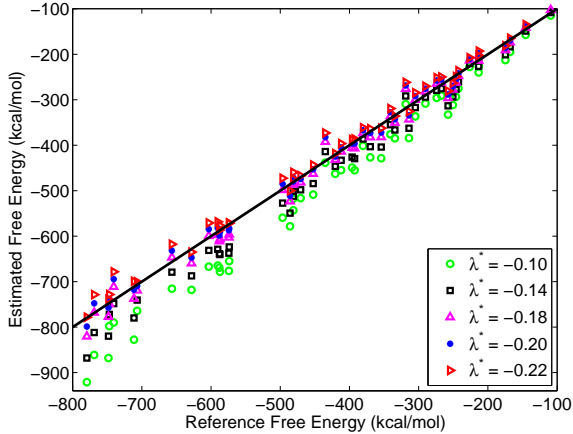}}
\caption{Dependence of modified BIBEE model accuracy on the eigenvalue
  approximation, taken over 50 structures from the test set of Feig
  \textit{et al.}~\cite{Feig04_GB_vs_PB} using CHARMM22 radii and
  charges and dielectric constants $\epsilon_1=4$ and $\epsilon_2=80$.
}\protect\label{fig:vary-lambda}
\end{figure}

As shown in Section~\ref{sec:analysis}, the BIBEE/P eigenvalue
approximation $\lambda=0$ is exact in the limit as the multipole order
goes to infinity.  An alternative modified BIBEE might employ instead
the GB$\epsilon$ strategy of choosing an approximation suitable for
the dipole and quadrupole; like BIBEE/M with $\lambda = 0$, this
strategy also leads to a representation that is the sum of separable
terms.  For the electric-field operator on the sphere, these
eigenvalues are $\lambda = -1/6$ and $\lambda = -1/10$; we therefore
tested a range of $\lambda$ from -0.10 to -0.22 on 50 of the proteins
from the Feig \textit{et al.} test set; we used the larger magnitude
limits to account for the larger dipole eigenvalues of ellipsoidal
geometries~\cite{Ritter95}. In fact, the simple parameter sweep
illustrated that $\lambda = -0.20$ offered approximately the best
accuracy (Figure~\ref{fig:vary-lambda}), with a
root-mean-square-deviation (RMSD) of 18.1~kcal/mol, corresponding to a
mean deviation of 3.6\%; Table~1 contains the corresponding results
for different $\lambda$.  Future work should examine the possibility
of identifying the molecule's approximate best-fit ellipsoid,
following the recent ALPB model of Sigalov, Onufriev \textit{et
  al.}~\cite{Sigalov06}; such an approach might provide a rigorous,
parameter-free model as opposed to the one-parameter approach here.
\begin{table*}
  \centering
  \protect\label{table:vary-lambda}
  \caption{Dependence of modified BIBEE model accuracy on the
    eigenvalue approximation, taken over 50 structures from the test
    set of Feig \textit{et al.}~\cite{Feig04_GB_vs_PB}.  See main text
    for calculation details.}
  \begin{tabular}{c|cccccc}\hline\hline
    $\lambda$ & -0.10 & -0.14 & -0.16 & -0.18 & -0.20 & -0.22 \\ \hline
    RMSD (kcal/mol) & 67.4 & 40.5 & 29.4 & 21.0 & 18.1 & 21.9 \\
    Mean deviation (\%) & 12.5 & 7.4 & 5.3 & 4.1 & 3.6 & 4.3 \\
    \end{tabular}
\end{table*}

\begin{figure}[ht!]
\centering
\resizebox{3.0in}{!}{\includegraphics{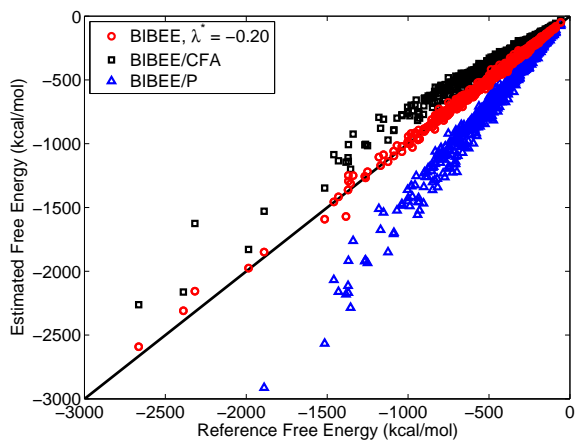}}
\caption{Electrostatic solvation free energies for all of the proteins
  from the test set of Feig \textit{et al.}~\cite{Feig04_GB_vs_PB},
  using CHARMM22 radii and charges and dielectric constants
  $\epsilon_1=4$ and $\epsilon_2=80$, the original BIBEE/CFA and
  BIBEE/P models, as well as the modified BIBEE with the eigenvalue
  approximation $\lambda = -0.20$.}\protect\label{fig:improved-BIBEE}
\end{figure}

\begin{figure}[ht!]
\centering
\resizebox{3.0in}{!}{\includegraphics{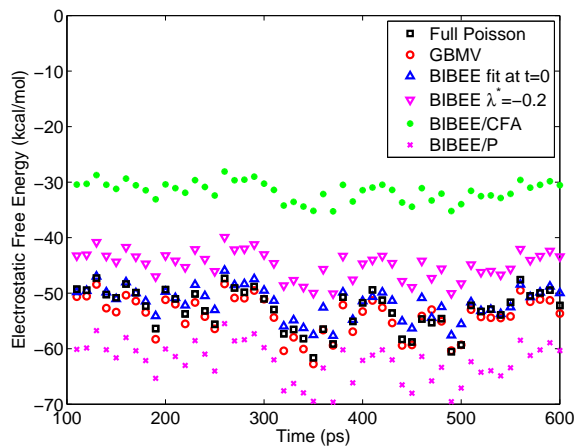}}
\caption{Electrostatic solvation free energies for 50 structures of
  met-enkephalin taken from an all-atom molecular dynamics simulation
  in explicit water.  The GBMV results were calculated using
  CHARMM~\cite{Brooks83} and the remainder were calculated using
  $\epsilon_1 = 1$, $\epsilon_2=80$, and CHARMM22 radii and charges.
  See text for other calculation details.}\protect\label{fig:met}
\end{figure}

We then used the single fit parameter $\lambda = -0.20$ to estimate
the electrostatic solvation free energies for the 610 proteins in the
Feig \textit{et al.} test set (Figure~\ref{fig:improved-BIBEE}); the
RMSD of the modified approach was 22.9~kcal/mol (3.2\% mean
deviation), illustrating that a simple one-parameter model is capable
of excellent accuracy on a wide range of protein sizes and shapes.
For a simple example of the model's capability to capture energetic
differences as a function of conformation, and also to illustrate the
model's performance compared to a modern GB theory, we employ 50
structures of the peptide met-enkephalin taken from all-atom molecular
dynamics in explicit solvent, which were generated in earlier
work~\cite{Bardhan09_bounds}.  Figure~\ref{fig:met} contains plots of
the numerically calculated free energies, as well as those computed
using the GBMV module of CHARMM~\cite{Brooks83}.  The $\lambda =
-0.20$ approximation is clearly a significant improvement over the
original BIBEE/CFA and BIBEE/P estimates, but further from the actual
numerical result than GBMV; the RMSD for GBMV is 1.41~kcal/mol.
However, if one first determines $\lambda$ so that the modified BIBEE
matches the numerical result at the first snapshot (labeled $t = 0$),
then the modified BIBEE is competitive with GBMV in accuracy
(1.88~kcal/mol).  This suggests that pre-computation of appropriate
fitting parameters, when rigorous approaches are found for their
derivation, may enable BIBEE methods to be suitable for calculating
averages as found in MM/PBSA computations\cite{Carrascal10}.

%\subsection{Simple Ansatz for the Sphere}\label{subsec:simple-ansatz}
%
%The exact solution (\ref{eq:BnmDef}) exhibits a non-separable
%representation of the dielectric constants and the mode $n$, whereas
%the BIBEE models possess the separable representations
%(\ref{eq:BnmCFADef}) and (\ref{eq:BnmPDef}).  Considering the
%separability discussion previously, it would seem that one might be
%able to design a fast and accurate model simply by finding functions
%$f(n)$ and $g(\epsilon_1,\epsilon_2)$ to match the exact solution as
%closely as possible; that is, we considered the possibility that a
%separable relationship similar to (\ref{eq:BnmCFADef}) could be
%converted back to a modified boundary condition and thus to a simple
%and accurate electrostatic model.  A simple example shows that this
%is unfortunately not possible:  the separable representation
%\begin{equation}
%  B^{S}_{nm} = \frac{1}{n+1} \frac{1}{\epsilon_2} \gamma_{nm}
%\end{equation}
%is nothing more than the limit of large $\epsilon_2$ shown in
%Eq.~\ref{eq:LargeEpsilonDiff}. For biomolecular solvation problems
%with $\epsilon_2/\epsilon_1 \cong 40$, this approximation shows
%excellent accuracy, always within a few percent for each multipole
%coefficient (Figure~\ref{fig:simple-ansatz}).%

%This relation does not convert in a simple way to a boundary
%condition:
%  

\section{Discussion}\label{sec:discussion}

In this paper, we have derived a complete analytical characterization
of the BIBEE approximation of Poisson electrostatics for charges in
spherical cavities.  The simplified analysis clarifies several
features of the BIBEE model, highlights an important commonality
between Generalized Born models and BIBEE, and suggests an immediate
correction scheme, BIBEE/M, with improved accuracy.  Future work will
focus on developing further mathematically rigorous improvements with
clear physical interpretation.  We emphasize that the BIBEE/M model
possesses the same attractive characteristics as the earlier BIBEE
methods, including the excellent preservation of reaction-potential
eigenfunctions for non-trivial geometries.

As may be expected from the initial derivation of the BIBEE
model~\cite{Bardhan08_BIBEE}, the BIBEE and GB methods share a key
feature in their approximations: both approaches estimate the
electrostatic solvation free energy using separation of geometric and
material variables.  In both approaches, the approximate free energy
is calculated as the product of two quantities, one purely a function
of the dielectric constants, and the other quantity, which is solely a
function of the geometry: dielectric boundary and charge distribution.
For the initial GB/CFA models, this quantity combined the effective
Born radii and the Still equation; for the BIBEE models, it relates to
the Coulomb potential operators.  Deviations from separability have
been recognized and analyzed in detail by Onufriev and
collaborators~\cite{Sigalov05,Sigalov06}.  In our work, we are
exploring how far a purely mathematical framework may go to develop an
acceptably accurate approximation to the continuum electrostatic
problem, that is, before one invokes approximations specific to
biomolecular simulations, such as the Still equation.

We have shown that for spherical geometries, the eigenfunctions of the
approximate reaction-potential operator are exact, a result that can
easily be shown to hold also for planar half-space problems.
Remarkably, the proof for spheres does not rely on the spectral
properties of the boundary-integral operators employed in proving that
BIBEE/CFA is a rigorous upper bound to the actual Poisson free
energy~\cite{Bardhan09_bounds}.  The eigenvector analysis can and
should be analyzed from this perspective also, which may furnish
additional insights into the BIBEE approximation.  In particular, for
the sphere it can be shown that the single-layer integral operator and
the normal electric field operator are related by a simple scale
factor; thus, for the sphere the operators share eigenvectors, and the
normal electric field is symmetric---which is not true for general
surfaces.  Further rigorous improvements may be possible for
nonspherical problems if operator analysis reveals the relative
importance of these properties in determining the accuracy of the
BIBEE eigenvectors.

The BIBEE/M analytical correction for the dominant mode described in
Section~\ref{sec:analysis} is not as easily extended to higher modes,
unfortunately, because the surface electric fields for the higher
modes are geometry dependent.  Using a nonzero $\lambda$ for the other
modes provides good accuracy on a range of problems, and if fit for a
particular protein can rival GBMV on certain tests, but it is
naturally preferable to design approximations that can be
systematically improved.  In future work, we will explore how the
dominant eigenvectors and eigenvalues of the integral operator may be
approximated to provide improved accuracy.

\section*{Acknowledgments}
We thank L. Greengard, M. Anitescu, and D. Gillespie for useful
discussions, B. Roux for the use of {\sc CHARMM}, and B. Tidor for the
use of the ICE (Integrated Continuum Electrostatics) software library.
The work of MGK was supported in part by the U.S. Army Research
Laboratory and the U.S. Army Research Office under contract/grant
number W911NF-09-0488.

%\bibliographystyle{plain}
%\bibliography{sphbib}

\end{document}